\documentclass[journal,a4paper,10pt]{IEEEtran}
\IEEEoverridecommandlockouts
\usepackage{times}
\usepackage{cite}
\usepackage{flushend}
\usepackage[dvips]{color}
\usepackage{epsf}
\usepackage{epsfig}
\usepackage{graphicx}
\usepackage{graphics}
\usepackage{epstopdf}
\usepackage{amsmath}
\usepackage{amssymb}
\usepackage{amsthm}
\usepackage{amsxtra}
\usepackage{algorithm}
\usepackage{algorithmic} 	
\usepackage{enumerate}
\usepackage{multirow}
\usepackage{amssymb}
\usepackage{multirow}
\usepackage{tikz}
\usepackage{fancyhdr}
\usepackage{rotating} 
\usepackage{graphicx} 

\usepackage{url}

\makeatother

\ifodd 1
\newcommand{\rev}[1]{{\color{blue} #1}} 
\newcommand{\com}[1]{\textbf{\color{blue} #1}}
\newcommand{\response}[1]{\textbf{\color{magenta} (RESPONSE: #1)}} 
\else
\newcommand{\rev}[1]{}
\newcommand{\com}[1]{}
\newcommand{\comg}[1]{}
\newcommand{\response}[1]{}
\fi

\begin{document}
\title{Customer Engagement Plans for Peak Load Reduction in Residential Smart Grids}
\author{Naveed Ul Hassan,~\IEEEmembership{Member,~IEEE,} Yawar Ismail Khalid, ~Chau~Yuen,~\IEEEmembership{Senior Member,~IEEE,} and Wayes~Tushar,~\IEEEmembership{Member,~IEEE}
\thanks{N. U. Hassan and Y. I. Khalid are with the Electrical Engineering Department at the Lahore University of Management Sciences, Lahore, Pakistan 54792. (Email: naveed.hassan@yahoo.com, yawar89@gmail.com).}
\thanks{W. Tushar and C. Yuen are with the Engineering Product Development at the Singapore University of Technology and Design (SUTD),  8 Somapah Road, Singapore 487372. (Email: \{wayes\_tushar, yuenchau\}@sutd.edu.sg).}
\thanks{This research is partly supported by Lahore University of Management Sciences (LUMS) Research Startup Grant and the Energy Innovation Research Program (EIRP) Singapore NRF2012EWT-EIRP002-045.}
}
\IEEEoverridecommandlockouts
\maketitle
\thispagestyle{fancy}
\begin{abstract}
In this paper, we propose and study the effectiveness of customer engagement plans that clearly specify the amount of intervention in customer's load settings by the grid operator for peak load reduction. We suggest two different types of plans, including Constant Deviation Plans (CDPs) and Proportional Deviation Plans (PDPs). We define an adjustable reference temperature for both CDPs and PDPs to limit the output temperature of each thermostat load and to control the number of devices eligible to participate in Demand Response Program (DRP). We model thermostat loads as power throttling devices and design algorithms to evaluate the impact of power throttling states and plan parameters on peak load reduction. Based on the simulation results, we recommend PDPs to the customers of a residential community with variable thermostat set point preferences, while CDPs are suitable for customers with similar thermostat set point preferences. If thermostat loads have multiple power throttling states, customer engagement plans with less temperature deviations from thermostat set points are recommended. Contrary to classical ON/OFF control, higher temperature deviations are required to achieve similar amount of peak load reduction. Several other interesting tradeoffs and useful guidelines for designing mutually beneficial incentives for both the grid operator and customers can also be identified.

\end{abstract}
 \setcounter{page}{1}

\section*{Nomenclature}
\addcontentsline{toc}{section}{Nomenclature}

\begin{IEEEdescription}[\IEEEsetlabelwidth{$a_1,~~~,b_1$}]
\item[$K$] Number of operable states of thermostat loads.
\item[$K_i^j$] Number of operable states of thermostat load $i$ of customer $j$.
\item[$k$] Index of each operable state of thermostat load.
\item[$J$] Number of customers in the residential community.
\item[$I$] Set of flexible loads for which customer engagement plans are defined.
\item[$I_T$] Set of thermostat loads for which customer engagement plans are defined.
\item[$I_S$] Set of shiftable loads for which customer engagement plans are defined.
\item[$I_{Tc}$] Set of thermostat loads used for cooling for which customer engagement plans are defined.
\item[$I_{Th}$] Set of thermostat loads used for heating for which customer engagement plans are defined.
\item[$I^j$] Set of flexible loads owned by customer $j$.
\item[$I_T^j$] Set of thermostat loads owned by customer $j$.
\item[$I_S^j$] Set of shiftable loads owned by customer $j$.
\item[$I_{Tc}^j$] Set of thermostat loads used for cooling owned by customer $j$.
\item[$I_{Th}^j$] Set of thermostat loads used for heating owned by customer $j$.
\item[$i$] Each flexible load index.
\item[$j$] Each customer index.
\item[$n$] Index of each local peak in Algorithm 2.
\item[$\hat{\theta}_{j,i}$] Preference of customer $j$ for the set point of thermostat load $i$.
\item[${\theta}^\text{ref}_{i}$] Reference temperature for thermostat load $i$.
\item[$\triangle{\theta}_{i}^\text{max}$] Constant value representing the maximum temperature deviation for thermostat load $i$.
\item[$\triangle{\theta}_{j,i}$] Inconvenience severity experienced by customer $j$ for thermostat load $i$.
\item[$\theta_{j,i}(t)$] Output temperature of thermostat load $i$ of customer $j$.
\item[$\theta_{j,\text{AC}}^{k}(t)$] Output temperature of AC of customer $j$, which is operated in state $k$, at time $t$.
\item[$\theta_{j,\text{WH}}^{k}(t)$] Output temperature of WH of customer $j$, which is operated in state $k$, at time $t$.
\item[$\bar{\theta}_{j,i}(t)$] Output temperature of thermostat load $i$ of customer $j$ at time $t$ during Algorithm 2 computations.
\item[$\theta_\text{AC}^\text{ave}$] Average temperature deviation of AC loads over the demanded intervals in the simulations.
\item[$\theta_\text{WH}^\text{ave}$] Average temperature deviation of WH loads over the demanded intervals in the simulations.
\item[$\theta_\text{inlet}$] Temperature of the inlet water of WH.
\item[$\theta_a(t)$] Room temperature at $t$.
\item[$fr(t)$] Rate of water flow in WH.
\item[$V_\text{tank}$] Volume of the water tank of WH.
\item[$A_\text{tank}$] Tank surface area of WH.
\item[$R_\text{tank}$] Heat resistance of the water tank of WH.
\item[$G_j(t)$] Heat gain rate of the house of customer $j$ at time $t$.
\item[$T$] Number of intervals in the considered time duration.
\item[$t$] Index of each time interval.
\item[$\triangle t$] Time duration of each interval.
\item[$\triangle s_i^\text{max}$] Maximum scheduling delay of each shiftable load $i$.
\item[$\Delta d_{i}^\text{max}$] Maximum inconvenience duration of each thermostat load $i$.
\item[$t_{j,i}^\text{start}$] Actual start time of shiftable load $i$ of customer $j$.
\item[$t_{j,i}^\text{pref}$] Preferred start time of shiftable load $i$ of customer $j$.
\item[$t_i^\text{dur}$] Maximum number of time slots a thermostat load $i$ can deny operation at its full rated power.
\item[$t_{j,i}^\text{max-n}$] Time index corresponding to $n$-th largest local peak in the demanded operation interval of thermostat load $i$ of customer $j$ in Algorithm~2.
\item[$\mathbf{t}_{j,i}^\text{peak}$] Time index vector containing local peaks in Algorithm~2.
\item[$\mathbf{b}_{j,i}$] Set of time indexes during which customer $j$ demands thermostat load $i$.
\item[$\beta_i$] Scaling factor for thermostat load $i$.
\item[$\mathbf{w}_{j,i}$] Demand status vector of flexible device $i$ of customer $j$.
\item[$w_{j,i}(t)$] Entry, which is a binary variable, of the demand status vector at $t$.
\item[$p_{j,i}^r$] Rated power of flexible load $i$ of customer $j$.
\item[$l$] Index to denote scheduling delay in terms of number of time slots in Algorithm~\ref{greedy:algo}.
\item[$\mathbf{p}_{j,i}$] Vector of demanded power requirement by flexible load $i$ of customer $j$.
\item[$p_{j,i}(t)$] Demanded power consumption requirement of load $i$ of customer $j$ at time $t$.
\item[$\mathbf{p}_{j,i}^l$] Demanded power requirement vector of shiftable load $i$ of customer $j$ when delayed by $l$ time slots.
\item[$\mathbf{C}_{j,i}$] A matrix of binary variables of order $T\times K_i^j$.
\item[$c_{j,i}^k(t)$] Binary variable indicating the operational state of thermostat load $i$ of customer $j$ in state $k$ at time $t$.
\item[$\mathbf{c}_{j,i}(t)$] $t$-th row vector of matrix $\mathbf{C}_{j,i}$.
\item[$\mathbf{c}_{j,i}^k$] $k$-th column vector of matrix $\mathbf{C}_{j,i}$.
\item[$\alpha$] Energy required for a unit degree rise in room temperature.
\item[$\mathbf{1}_{(\cdot)}$] $(\cdot)\times 1$ vector of all 1's.
\item[$\mathbf{e}_{K_i^j}^\tau$] Vector of all 1's except a $0$ at $K_i^j$-th position. 
\item[$\mathbf{x}$] A given aggregated load profile vector of residential community.
\item[$x(t)$] Aggregated load demand at time $t$.
\item[$\hat{\mathbf{x}}$] Aggregated load profile obtained by solving sub-problem 1. 
\item[$\hat{\mathbf{x}}_j$] Aggregated load profile obtained from the customer $j$ in Algorithm~\ref{greedy:algo}.
\item[$\tilde{\mathbf{x}}$] Final aggregated output load profile.
\item[$\mathcal{F}(\cdot)$] A circular shift operator.
\item[$m$] Particular order of shiftable devices.
\item[$m^*$] Optimal order of shiftable devices.
\item[$\mathbf{y}^l$] Aggregated load profile in Algorithm~\ref{greedy:algo} when shiftable load $i$ of customer $j$ is delayed by $l$ time slots.
\item[${\phi}_l$] Peak value of $\mathbf{y}^l$ in Algorithm~\ref{greedy:algo}.
\item[$l_i^*$] Optimal scheduling delay for shiftable load $i$ determined by Algorithm~\ref{greedy:algo}.
\item[$\mathbf{y}$] Output of Algorithm~\ref{greedy:algo} (aggregated load profile).
\item[$\mathbf{y}_m$] Output of Algorithm~\ref{greedy:algo} for $m$-th particular order (aggregated load profile).
\item[$\hat{\phi}_m$] Peak value of $\mathbf{y}_m$ (for shiftable device order $m$).
\item[$q_{j,i}^k(t)$] Power consumption of thermostat load $i$ of customer $j$ while operating in state $k$ at time $t$.
\item[$\mathbf{Q}_{j,i}$] A matrix of order $T\times K_i^j$, which has the entry of $q_{j,i}^k(t),~\forall j, k, i\in I_T$.
\item[$\hat{Z}_{j,\text{AC}}^k$] Cooling capacity of customer $j$'s AC operating in state $k$ at time $t$.
\item[$EER$] Energy efficiency ratio of the AC.
\item[$\mathbf{h}$] Output of Algorithm~2 (aggregated load profile).
\item[$N_\text{AC}$] Number of eligible AC loads in the simulations.
\item[$N_\text{AC}$] Number of eligible WH loads in the simulations.
\end{IEEEdescription}

\section{Introduction}\label{sec:introduction}
Demand Response Program (DRP) can be used to reduce cost and improve efficiency of power grids by engaging customers and modifying their power consumption pattern~\cite{Liu-JSTSP:2014,Tushar-TIE:2014,Tushar-ISGT:2014,bochai-2014,Tushar-globecom:2014,Naveed-ISGT1:2013,Naveed-ISGT2:2013}. Current smart metering and bi-directional communication technologies, allow the inclusion of industrial as well as residential consumers in DRPs \cite{DRM_1, DRM_2, DRM_3,Liu:TIE:2014,Wayes-J-TSG:2012,Tushar-TSG:2013}. Customer engagement for peak load reduction can be achieved through motivating customers by either time-based DRP or incentive-based DRP \cite{survey,ref2}. In time-based DRP, electricity prices are dynamically varied at different times and the customers can modify their electricity consumption in response to the change of price. In incentive-based DRP, fixed or time-varying payments are offered to the customers under specific constraints. In general, residential customers are not very responsive to DRPs due to the uncertainty in their electricity bills and the lack of clarity about the resulting inconvenience that they might experience \cite{nordic}. The participation of residential customers in DRP, however, is extremely important to address the predictable and non-predictable supply and demand variations in order to reduce the electricity generation cost \cite{elec_price1, elec_price3}. 

The increased penetration and usage of Air Conditioner (AC) and Water Heater (WH) loads in residential sector are the main reasons for increasing the peak load demand on the grid \cite{koomey,new_aus,hongkong}. In \cite{wh_new}, the authors propose a direct load control algorithm to investigate the potential of WH loads in providing load-balancing service, while in \cite{ref3}, the authors evaluate the potential of Heating, Ventilation and Air-Conditioning (HVAC) loads for providing regulation services. These two papers suggest that the aggregated regulation service  provided by the WH and HVAC loads can become one of the major sources of revenue for the grid. The authors in \cite{hvac_aa} propose a controller for thermostatically controlled loads, which can be used for peak shaving and load shifting by managing the aggregated HVAC load shapes. The authors in \cite{hvac_bb} propose HVAC control algorithm to regulate the indoor air temperature inside a defined dead band by using an ON/OFF power control. However, the algorithm does not consider the retail price of electricity in the HVAC control, and therefore may fail to curtail load during peak price period. 

In \cite{hvac_cc}, a price responsive control strategy is proposed for HVACs to reduce peak load. The controller changes the thermostat set point of HVAC loads depending on the electricity retail price published in every 15 minutes. The authors demonstrate a significant peak load reduction with a modest variation in thermal comfort. In \cite{large_1}, the authors consider the problem of managing large number thermostat appliances with ON/OFF power control for peak load reduction. They discuss a centralized, model predictive approach and a distributed structure with a randomized dispatch strategy. 

Several authors have also designed algorithms that may help the grid operator to reduce peak demand by controlling and scheduling common household appliances. For instance, \cite{rev_ref} proposes an optimal and automatic residential load commitment framework, in which the households achieve the minimum payment by responding to the time-varying prices offered by the grid. Similarly, a Stackelberg game is studied in \cite{stackel_1} to maximize both the revenue of the grid operator and the payoff to each user by determining the optimal electricity price and the optimal price consumption. In \cite{shinwari_1}, the authors use a water-filling based scheduling algorithm, which does not require any communication between scheduling loads, to obtain a flat demand curve. The authors also explore the possible errors in demand forecast and potential incentives for customer participation. A genetic algorithm based energy resource scheduling technique, which includes day-ahead, hour-ahead, and five minutes ahead scheduling is proposed for smart grid in \cite{mnlp_1}. Finally, a heuristic method called Signaled Particle Swarm Optimization is proposed in \cite{swarm_1} for distributed energy resource scheduling in smart grids.

These DRP algorithms certainly present feasible solutions to address the problem of increased peak load on the grid. However, designing schemes, which can attract the interest of significantly large number of customers to participate in DRP remains a major challenge~\cite{nordic,elec_price1, elec_price3}. The reluctance to participate by the residential customers is due to the lack of clarity in specifying key information such as the number of times customers would be called upon to participate, the range of temperature variation of AC and WH loads, and the financial benefits associated with such participation. In recent years, companies like Idaho Power have introduced direct load control methods, such as AC Cool Credit program, to switch OFF the AC loads of their residential customers \cite{idaho}, where the duration of OFF time is set according to a pre-determined agreement. Such agreement has been reported to be very effective in considerably reducing the peak load. However, the resulting room temperature deviations during the OFF intervals are not specified in the agreement, which can result in excessively high inconvenience for some customers. 

In this paper, we propose customer engagement plans that clearly specifies all the key inconvenience parameters for shiftable and thermostat loads, and thus may encourage the customers to pick plans according to their behavioral and financial requirements\footnote{Designing actual financial rewards or incentives for customers is out of scope of this paper.}. For each shiftable load, maximum scheduling delay is specified in the plan. For each thermostat load, maximum temperature deviation from the thermostat set point and the maximum time duration during which the actual temperature deviates from the thermostat set point are clearly defined\footnote{Customers generally cannot understand the amount of intervention by specifying the energy consumption limits on their loads or by announcing the dynamic price on energy consumption. It is much easier for them to comprehend the inconvenience if specified in terms of scheduling delays and temperature deviations of loads.}. It is important to note that this work is an extension of our previous work in \cite{smart_grid_ref}. In \cite{smart_grid_ref}, we proposed demand response management plans for AC loads. We modeled the AC load as a power throttling thermostat device and specified the thermostat set point, temperature deviation and inconvenience duration for the customers. We showed that a power throttling thermostat device can operate in $K\geq 2$ states, where $K$ is the number of possible power states. Note that in a 2-state model, thermostat loads can only be turned ON/OFF for DRP, whereby they can operate in $K$ different states for $K>2$. For example,  in a 3-state model, the thermostat load can be turned OFF, operated at $50\%$ of the rated power or operated at full rated power. In \cite{smart_grid_ref}, we determined the effectiveness of such plans and studied the impact of temperature deviation, time duration of inconvenience, and the impact of increasing the power throttling states on peak load reduction. However, the inclusion of thermostat set point as a plan parameter in \cite{smart_grid_ref}, requires all the customers to adjust their thermostat set point to the same specified value and thus experience additional temperature deviation for the time duration as laid out in the plan. Such customer engagement plans, therefore, may have  practical limitations and fairness concerns.

To this end, we extend the work in \cite{smart_grid_ref} in this paper to include multiple shiftable and thermostat loads. In this context, we propose two new types of customer engagement plans: the Constant Deviation Plan (CDP) and Proportional Deviation Plan (PDP), which allow customers to have different thermostat set points. We model thermostat loads as power throttling devices such as in \cite{smart_grid_ref}. However, in contrast to \cite{smart_grid_ref}, the model adopted in this paper is quite generic and can allow different customers to have different number of loads including thermostat loads with different number of power throttling states. We propose algorithms that enable us to identify mutually beneficial plans for both the grid operator and the customers with power throttling thermostats and shiftable loads through the introduced customer engagement plans, i.e., CDPs and PDPs. Through MATLAB simulations, we study the impact of the number of power throttling states, magnitude of temperature deviations, and scheduling delays on the peak load reduction. The impact of scheduling delays in the presence of thermostat loads with multiple power throttling states is investigated, and certain observations and comparisons of CDP and PDP type plans are made. The impact of CDP and PDP parameters in controlling and determining the number of eligible thermostat loads and average temperature deviations are also investigated. Some useful design guidelines for CDP and PDP plans are provided, which could be helpful in designing the financial incentives for participating customers. Further, several new and interesting research directions are identified.

The rest of the paper is organized as follows; the customer load model, residential community model and customer engagement plans are explained in Section II, the optimization problem for peak load reduction and the algorithmic solutions are presented in Section III, simulation results are given in Section IV, while the paper is concluded and future research directions are identified in Section V.

\section{Customer Load, Residential Community Models and Customer Engagement Plans}
\subsection{Customer Load Model}
In this paper, we group the appliances, also called loads or devices, in a home into two categories: essential appliances, and flexible appliances.  The power required for all the entire essential appliances serves as a base load, which always has to be provided by the grid operator. We further classify flexible loads into shiftable and thermostat loads\footnote{Other categories of flexible loads e.g., electrical vehicles, which require an unfixed amount of power and operation duration depending on the customer usage are also possible. Such load categories are not considered in this paper. However, our framework can be easily extended in future to accommodate electrical vehicles or any other new category of flexible loads. }. We assume that the shiftable loads, such as Dishwashers (DWs) or Clothes Dryer (CD), require a constant power draw for a specified duration. We further assume that neither the power draw nor the operational duration of shiftable loads can be reduced. The customer has a preferred operation interval for each shiftable load that best suits his/her behavioral requirements. Scheduling delays in the operation of shiftable loads, can cause inconvenience to customers. 

Thermostat loads such as AC and WH, on the other hand, have to maintain an output temperature that matches the customer-defined thermostat set point, and any deviation from these set points cause inconvenience. Temperature deviation can be controlled in two different ways: 1) by re-adjusting the thermostat set point; or 2) by re-adjusting the power consumption of the device. During the re-adjustment of set point, a smart thermostat automatically adjusts the thermostat settings to control temperature deviations. One such example is Google Nest, which can be used to re-adjust the thermostat set point of AC loads \cite{googlenest}. Conversely, power consumption can be re-adjusted by providing a smart electrical interface with each device which enables it to respond to the control signals received from a controller. For example, Idaho Power controls the temperature deviations of AC loads by adjusting the number of ON/OFF switching frequency. Thermostat re-adjustment method requires mathematical models (for each thermostat load) to compute the resulting power consumption for some specified temperature deviation, while power re-adjustment method requires models to compute the resulting temperature deviations for specified power consumption. 

In this paper, we consider thermostat loads that are capable of operating in $K\geq 2$ power consumption states. We have also adopted the power re-adjustment method to control the temperature deviations of thermostat appliances \cite{hvac_bb,hvac_cc,large_1,idaho,g14}. This method allows us to determine the exact amount of power consumption by the appliances and using the thermal models readily available in the literature \cite{g14}, we can determine the output temperature (e.g., room temperature or hot water temperature). Moreover, the thermal models can be easily extended to include multiple power throttling states. The AS/NZS 4755 standard (jointly adopted by the Australian and New Zealand Governments) \cite{aus_stand} for thermostat loads being manufactured and sold in Australia and New Zealand also mandates physical/electrical interface as well as mandatory and optional modes, which permit these loads to operate in different power consumption states. Variable frequency drives and variable speed drives are generally used to throttle power of thermostat loads between different states.

\subsection{Residential Community Model} 
We model a residential community comprising of $J$ customers (also called homes, users or consumers). We consider the aggregated power consumption profile of the community for a duration of 24 hours, and divide the total duration into $T$ equal intervals. Each interval $\Delta t$ comprises of $\Delta t = \frac{24\times 60}{T}$ minutes. Each customer has a base load, a set of thermostat loads and a set of shiftable loads. We further differentiate thermostat loads into two types depending on whether the load is used for cooling application or heating application. For example, in summer AC is used to cool the room, while WH is used to heat water. In our paper, we assume that the grid operator identifies a set of flexible devices denoted by $I$, further categorized into $I_S$ shiftable and $I_T$ thermostat loads, which it wants to control. Let $I_{Tc}$ and $I_{Th}$ respectively denote the set of thermostat loads used for cooling and heating applications ($I_T=I_{Tc} \cup I_{Th}$). The grid operator will then propose customer engagement plans only for these flexible devices. Let $I^j \subseteq I$, $I_S^j \subseteq I_S$ and $I_T^j \subseteq I_T$ respectively denote the set of flexible, shiftable and thermostat loads of customer $j$. Similarly, let $I_T^j=I_{Tc}^j \cup I_{Th}^j$, where, $I_{Tc}^j \subseteq I_{Tc}$ and $I_{Th}^j \subseteq I_{Th}$ denotes the set of thermostat loads of customer $j$ used respectively for cooling and heating applications. All the remaining devices of customer $j$ are treated as essential loads. For example, if the grid operator is only interested in controlling the AC load in a residential community and defines a customer engagement plan only for this load, then all the remaining loads such as CD or WH of all the customers will be treated as essential loads. This model, however, is flexible enough to allow different customers to have different number of flexible devices. This model also allows $I^j$ to be a strict subset of $I$, i.e., it is possible for some customers to own less flexible loads than the number of flexible devices $I$ for which the grid operator has defined customer engagement plans. In such a case, it is up to the grid operator to either allow such customers to participate in DRP with their available devices and offer them reduced financial incentives or does not allow them at all. In the rest of the paper, in order to have a generic model, we assume that a customer with $I^j \subset I$ is allowed by the grid operator to participate in the DRP (the problem on how the financial incentives are computed is out of scope of this paper).  

\begin{figure}[htb]
\centering
\includegraphics[width=0.45\textwidth,height=.25\textheight]{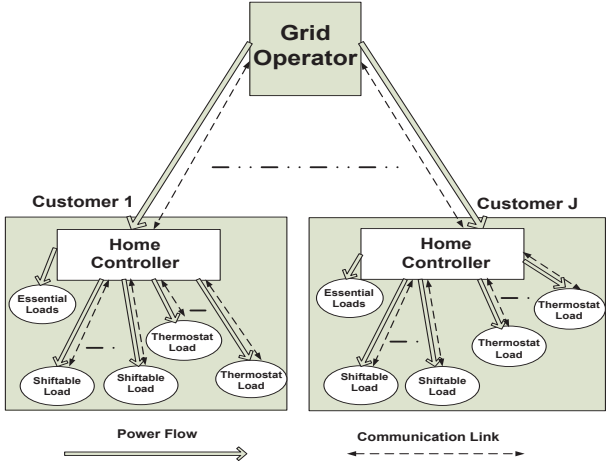}
\caption{System Model elaborating the implementation and communication requirements of our proposed methodology }
\label{fig:sysm}
\end{figure}

Each home is assumed to be equipped with a home controller as part of its advance metering infrastructure, which acts as an interface between the customer and the grid operator. There is a two-way communication link between the grid operator and each home controller. In our proposed method, the grid operator and the home controller only exchange the aggregated load profiles in order to minimize the communication bandwidth required and also to protect the privacy of individual customers. The home controller executes the algorithms (proposed in Sections III-A and III-B) in order to determine the scheduling time and power throttling states of its appliances according to the customer engagement plan. These values are stored at the home controller and the new aggregated load profile is communicated back to the grid. The home controller then schedules and operates its appliances accordingly. There is a two-way communication link between the home controller and every flexible load (shiftable or thermostat). The power flow is unidirectional i.e., from the grid operator to the home controller and appliances. The home controller generates appropriate control signals for each of its flexible device. These devices report back their status (ON/OFF status, power throttling state, temperature etc.) to the home controller. The implementation and communication requirements of our proposed methodology are further elaborated in Fig. \ref{fig:sysm}. 


\subsection{Customer Engagement Plans}
Customer engagement plans can be designed in one of the following ways: 
\begin{itemize}
	\item by defining the power consumption pattern for customers e.g., by announcing power consumption limits in peak load hours;
	\item by defining the inconvenience parameters for flexible loads of the customers.
\end{itemize}
Generally, it is difficult for customers to translate power consumption limits into scheduling delays and temperature deviations experienced by the loads, and easier for the grid operator to understand the amount of peak reduction and demand shaping. On the other hand, customer engagement plans that define the inconvenience parameters for the flexible loads can be easily understood by the customers, while their effectiveness for the grid operator is unclear. The second approach, however, is more promising to convince and encourage customer participation (because it provides clarity to the customers), which is therefore adopted in this paper. 

In our proposed customer engagement plans, the inconvenience related to any shiftable load $i\in I_S$ is defined in terms of scheduling delay from the preferred start time of the load. The plan defines $\Delta s_i^\text{max}$ as the maximum scheduling delay (in minutes) for each shiftable device $i$. The inconvenience related to any thermostat load $i \in I_T$ have two dimensions: 1) inconvenience duration: which is defined as the total time duration during the demanded interval the thermostat load is denied operation at its full rated power; and 2) inconvenience severity: which is defined as the temperature deviation from the desired thermostat set point. Maximum inconvenience duration $\Delta d_{i}^\text{max}$ (in minutes) for each thermostat load $i$ is specified in the plan. However, defining inconvenience severity is not straightforward since each customer has a different preference for thermostat set point (denoted by $\hat{\theta}_{j,i}$), which also impacts the experienced inconvenience and thermal comfort \cite{ashrae_stand,mac_per}. For example, if the outdoor temperature on a hot summer day is 82$^o$F, then a temperature deviation of 4$^o$F is more severe for a customer with AC thermostat set point at 76$^o$F compared to another customer with AC thermostat set point at 70$^o$F. We therefore propose two different types of plans: CDPs and PDPs. For each thermostat load $i$, both CDP and PDP define a reference temperature denoted by $\theta_i^\text{ref}$. 

In CDPs, a constant value representing the maximum temperature deviation for each thermostat load $i$, denoted by $\Delta \theta_{i}^\text{max}$, is announced. Accordingly, the inconvenience severity experienced by customer $j$ for thermostat load $i$ is computed as,
\begin{eqnarray}
\Delta \theta_{j,i}= \left\{ \begin{array}{ll}
 \left(\min(\theta_i^\text{ref}-\hat{\theta}_{j,i},\Delta \theta_{i}^\text{max})\right)^+ &\mbox{$\forall i\in I_{Tc}^j$}\\
  \left(\min(\hat{\theta}_{j,i}-\theta_i^\text{ref},\Delta \theta_{i}^\text{max})\right)^+ &\mbox{$\forall i\in I_{Th}^j$} 
       \end{array} \right. 
			\label{fixed_severity}
\end{eqnarray}
where, $x^+$ is defined by, $x^+=\max\{0,x\}$. In PDPs, a scaling factor $0 < \beta_i \leq 1$ is defined in the plan for each thermostat load $i$. The inconvenience severity experienced by customer $j$ for thermostat load $i$ is then computed as,
\begin{eqnarray}
\Delta \theta_{j,i}= \left\{ \begin{array}{ll}
 \beta_i \times (\theta_i^\text{ref}-\hat{\theta}_{j,i})^+ &\mbox{$\forall i\in I_{Tc}^j$}\\
 \beta_i \times (\hat{\theta}_{j,i}-\theta_i^\text{ref})^+ &\mbox{$\forall i\in I_{Th}^j$} 
       \end{array} \right. 
			\label{variable_severity}
\end{eqnarray}
To summarize; for each thermostat load $i$ following parameters are defined by the grid operator:
\begin{itemize}
	\item CDP: ($\Delta d_{i}^\text{max}$,$\Delta \theta_{i}^\text{max}$,$\theta_i^\text{ref}$) i.e., (Maximum inconvenience duration, Maximum temperature deviation, Reference temperature) and \eqref{fixed_severity};
	\item PDP: ($\Delta d_{i}^\text{max}$,$\beta_{i}$,$\theta_i^\text{ref}$) i.e., (Maximum inconvenience duration, Scaling factor, Reference temperature) and \eqref{variable_severity}.
\end{itemize}
As mentioned before, the reference temperatures in CDPs and PDPs bound the output temperature of thermostat loads, and can also be used to control the total number of loads experiencing temperature deviations. For example, if the thermostat set point of AC is below the reference temperature, then the output temperature for this customer cannot exceed the reference temperature. On the other hand, if the thermostat set point of AC is above the reference temperature, then such a customer will be allowed to operate at the preferred set point without any intervention. Generally, a high value of reference temperature for AC and a low value of reference temperature for WH will make a high number of AC and WH loads eligible for DRP. The values of reference temperatures can be decided by the grid operator depending on the local weather, geographical location, user preferences and the desired peak load reduction.

A sample CDP and PDP for a residential community comprising of AC, WH, CD and DW as flexible loads is given in Table \ref{tab:samp}. In this table, we also give an example of resulting temperature deviations of AC and WH loads of two customers with different thermostat set points to explain the difference between CDP and PDP and the role of reference temperatures (temperature deviations are computed using \eqref{fixed_severity} for CDP and \eqref{variable_severity} for PDP). Note that the thermostat set points of AC and WH loads of the customers in this example, are assumed to lie in a typical range associated with the thermal comfort of humans \cite{ashrae_stand,mac_per}. 

\begin{table*}[htb]
\renewcommand{\arraystretch}{1.2}
\renewcommand{\tabcolsep}{3pt}
\caption{Sample Customer Engagement Plans and an example of Resulting Temperature Deviations in CDP and PDP for two customers }
\centering
\begin{tabular} {|c|c| c| c| c|c| c| c| c|  |c|c| c| c| c|c| c| c|  }

\cline{1-17}
& \multicolumn{3}{ |c| }{AC} &  \multicolumn{3}{ c| }{WH} & CD & DW  & \multicolumn{2}{ |c| }{Customer-1: AC} & \multicolumn{2}{ |c| }{Customer-1: WH} & \multicolumn{2}{ |c| }{Customer-2: AC} & \multicolumn{2}{ |c| }{Customer-2: WH} \\ \hline

{\multirow{2}{*}{CDP}} & $\Delta d_\text{AC}^\text{max}$  & $\Delta \theta_\text{AC}^\text{max} $ & $\theta_\text{AC}^\text{ref}$ & $\Delta d_\text{WH}^\text{max}$ & $\Delta \theta_\text{WH}^\text{max} $  & $\theta_\text{WH}^\text{ref}$ &  $\Delta s_\text{CD}^\text{max} $ & $\Delta s_\text{DW}^\text{max} $ & $\hat{\theta}_{1,\text{AC}}$ & $\Delta \theta_{1,\text{AC}}$ & $\hat{\theta}_{2,\text{AC}}$ & $\Delta \theta_{2,\text{AC}}$ & $\hat{\theta}_{1,\text{WH}}$ & $\Delta \theta_{1,\text{WH}}$ & $\hat{\theta}_{2,\text{WH}}$ & $\Delta \theta_{2,\text{WH}}$ \\

 \cline{2-17}
& 30 min  & 4$^o$F & 80$^o$F & 30 min  & 8$^o$F & 108$^o$F & 60 min & 60 min & 70$^o$F & 4$^o$F & 76$^o$F & 4$^o$F & 114$^o$F & 6$^o$F & 104$^o$F & 0$^o$F \\ \hline \hline


{\multirow{2}{*}{PDP}} & $\Delta d_\text{AC}^\text{max}$  & $\beta_\text{AC} $ & $\theta_\text{AC}^\text{ref}$ & $\Delta d_\text{WH}^\text{max}$  & $\beta_\text{WH} $ & $\theta_\text{WH}^\text{ref}$ &  $\Delta s_\text{CD}^\text{max} $ & $\Delta s_\text{DW}^\text{max} $ & $\hat{\theta}_{1,\text{AC}}$ & $\Delta \theta_{1,\text{AC}}$ & $\hat{\theta}_{2,\text{AC}}$ & $\Delta \theta_{2,\text{AC}}$ & $\hat{\theta}_{1,\text{WH}}$ & $\Delta \theta_{1,\text{WH}}$ & $\hat{\theta}_{2,\text{WH}}$ & $\Delta \theta_{2,\text{WH}}$ \\

 \cline{2-17}
& 30 min & 0.6  & 80$^o$F & 30 min  & 0.8 & 108$^o$F & 60 min & 60 min & 70$^o$F & 6$^o$F & 76$^o$F & 2.4$^o$F & 114$^o$F & 4.8$^o$F & 104$^o$F & 0$^o$F \\ \hline

\end{tabular}
\label{tab:samp}
\end{table*}

The customers may subscribe to a plan on a daily, weekly or monthly basis. Similarly the grid operator can also change the set of its customer engagement plans on monthly, bi-monthly or seasonal basis (depending on the geography, weather patterns etc.). Such issues, however, are not discussed in this paper and are left as an interesting future work.

\section{Optimization Problem Formulation and Algorithm Development}
The objective of this paper is to determine the effectiveness of a given customer engagement plan (CDP or PDP) for peak load reduction. In this section, we mathematically formulate the optimization problem. Let $\textbf{w}_{j,i}=[ w_{j,i}(1),\ldots,w_{j,i}(T) ]^{\tau}$ denote the demand status vector of flexible device $i$ of user $j$ ($[\cdot]^{\tau}$ denotes the transpose operation). Each entry $w_{j,i}(t)$ of this vector is a binary variable: a `1' indicates that the device operation is demanded, while a `0' indicates that it is not demanded. Let $p_{j,i}^{r}$ denote the rated power of flexible load $i$ of customer $j$. Then $\textbf{p}_{j,i}=[p_{j,i}(1),\ldots,p_{j,i}(T)]^{\tau}=p_{j,i}^{r}\textbf{w}_{j,i}$ denotes the demanded power requirement vector of flexible load $i$ of customer $j$, while $p_{j,i}(t)$ is used to denote the demanded power consumption requirement of load $i$ of customer $j$ at time $t$. The operation of shiftable loads can be delayed from the preferred time interval. Let $t_{j,i}^\text{start}$ denote the actual start time of shiftable load $i$ of customer $j$, then according to the plan, we have the following constraint,
\begin{equation}
t_{j,i}^\text{start}-t_{j,i}^\text{pref} \leq \Delta s_{i}^\text{max}, \quad \forall j, i\in I_S^j
\label{shift_const}
\end{equation}
where, $t_{j,i}^\text{pref}$ denotes the preferred start time of shiftable load $i$ of user $j$. This constraint is sufficient to describe the terms of engagement as laid out in the plan for the shiftable loads. Also note that this constraint only allows shiftable loads to be delayed from their preferred start time (our framework can easily allow advance scheduling of shiftable loads, which however, is not considered in this paper). 

We assume that thermostat load $i$ of customer $j$ has the capability to operate in $K_i^j$ possible states. This model allows different customers to own thermostat loads with different number of operable states. For example, the AC of customer 1 can operate in only 2 states, while that of customer 2 can have 5 states. Let $\textbf{C}_{j,i}$ denote a $T\times K_i^j$ matrix of binary variables. The entries of this matrix are given by variables $c_{j,i}^k(t)$, which represent the operational state of thermostat load $i\in I_T^j$ of customer $j$ in state $k$ at time $t$. If thermostat load of customer $j$ is operated in state $k$ in time interval $t$, then $c_{j,i}^k(t)$ is set to `1'; otherwise to `0'. In the subsequent discussion, $\textbf{c}_{j,i}^k=[ c_{j,i}^k(1),\ldots,c_{j,i}^k(T) ]^{\tau}$ will be used to denote the $k$-th column vector, while $\textbf{c}_{j,i}(t)=[ c_{j,i}^1(t),\ldots,c_{j,i}^{K_i^j}(t) ]^{\tau}$ will be used to denote the $t$-th row vector of the matrix $\textbf{C}_{j,i}$. Since a thermostat load can operate in only one state in any given time interval $t$, we have the following constraint on its operation,
\begin{equation}
\textbf{C}_{j,i} \textbf{1}_{K_i^j}=\textbf{1}_T,\quad \forall j, i\in I_T^j
\label{oper_const1}
\end{equation}
where, $\textbf{1}_{K_i^j}$ is a $K_i^j\times 1$ vector of all 1's, while $\textbf{1}_T$ denotes a $T\times 1$ vector of all 1's.
Similarly, we can represent the constraint on the maximum inconvenience duration, i.e., $\Delta d_i^\text{max}$ of thermostat load $i$ mathematically as,
\begin{equation}
\textbf{e}_{K_i^j}^{\tau} \textbf{C}_{j,i}^{\tau} \textbf{w}_{j,i} \leq \Delta d_i^\text{max}, \quad \forall j, i\in I_T^j
\label{oper_const2}
\end{equation}
where, $\textbf{e}_{K_i^j}^{\tau}=[1\quad 1 \ldots 1 \quad 0]$ i.e., it is a vector of all 1's except a 0 in the $K_i^j$-th position. This constraint bounds the number of time slots in the demanded interval during which the operation of thermostat load is denied at its full rated power. The output temperature of a thermostat load at any time $t$ depends on its operational state. Let $\theta_{j,i}(\textbf{c}_{j,i}(t))$ denote the output temperature of thermostat load $i$ of customer $j$ at time $t$ (appropriate mathematical models are required to determine the output temperature as a function of power state of the load). During the time slots when the thermostat load is demanded by the customer, the difference in output temperature between any two consecutive intervals should be less than or equal to $\Delta \theta_{j,i}$ (computed using \eqref{fixed_severity} for CDP and \eqref{variable_severity} for PDP), which can be expressed in terms of the following constraint, 
\begin{equation}
w_{j,i}(t) \left|\theta_{j,i}(\textbf{c}_{j,i}(t)) - \hat{\theta}_{j,i} \right| \leq \Delta \theta_{j,i}, \: \forall j, t, i\in I_T^j
\label{temp_const}
\end{equation}

\begin{figure}[htb]
\centering
\includegraphics[width=0.46\textwidth,height=.08\textheight]{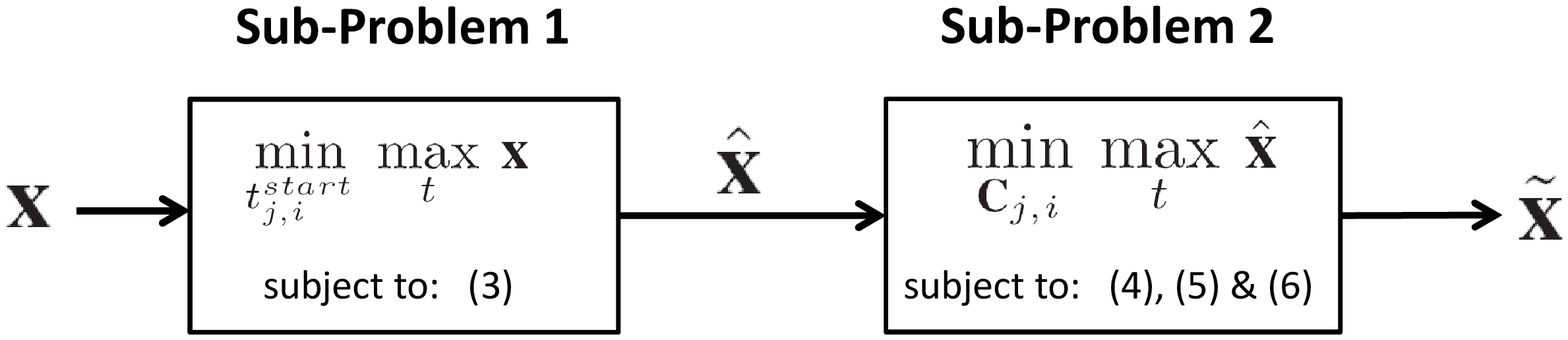}
\caption{Decomposition of the Optimization Problem according to Load Categories}
\label{fig:fig001}
\end{figure}

As stated before, our objective is to determine the effectiveness of customer engagement plan in terms of peak load reduction. Let $\textbf{x}=[x(1) \ldots x(T) ]^{\tau}$ denote the given aggregated load profile vector of the residential community, where, $x(t)=\sum_{j} \sum_{i} p_{j,i}(t)$. Then we have the following optimization problem, 
\begin{equation}
\min_{\textbf{C}_{j,i},t_{j,i}^\text{start}} \: \max_{t} \quad \textbf{x}
\label{obj:fun}
\end{equation}
\begin{equation}
\text{subject to constraints: \eqref{shift_const}, \eqref{oper_const1}, \eqref{oper_const2} and \eqref{temp_const} } \nonumber
\end{equation}
The objective function \eqref{obj:fun} seeks to minimize the maximum value of $\textbf{x}$. The optimal solution of this problem depends on the order in which customers are considered for optimization. Due to the combinatorial nature of the problem, it is NP-hard and therefore finding the optimal solution has exponential time complexity \cite{NP_ref1,NP_ref2,NP_ref3}. We develop a sub-optimal heuristic by decomposing the problem into two sub-problems based on the load categories, i.e., shiftable and thermostat loads as shown in Fig. \ref{fig:fig001}. We first minimize the peak load by re-scheduling the shiftable loads subject to constraint \eqref{shift_const}. The new aggregated load profile is denoted by $\hat{\textbf{x}}$. The peak of this load profile is further reduced by controlling the thermostat loads subject to constraints \eqref{oper_const1}, \eqref{oper_const2} and \eqref{temp_const} to obtain the final aggregated output load profile $\tilde{\textbf{x}}$. It should be noted that the order in which these steps are carried out can be interchanged but could result in a different amount of peak load reduction. 

Our sub-division of the problem and the subsequent algorithms based on the load categories provides flexibility and easily allows us to include or exclude some load category from optimization. For example, the grid operator can implement the framework without the shiftable load category or without the thermostat load category if required controls (interfaces) are not available. Moreover, if a third category of flexible loads (e.g., electrical vehicles) is available, then the framework can be easily extended by adding a third algorithm for the new category of flexible loads. Such flexibility can also be helpful in the actual implementation, where the proposed framework can be implemented in stages, as the interfaces and controls can be different for different load categories. Other alternatives, in which all the appliances of a customer are considered before moving on to the second customer are less flexible compared to our approach.

In Sections III-A and III-B, we develop distributed offline algorithms for the two sub-problems. We assume prior knowledge on the aggregated load profile of the residential community. This information can be obtained either from past power consumption patterns of the community during the same time period or it can be deduced using some load prediction/forecast models \cite{forecast1,forecast4}. 

\textbf{Remark}: It should be noted that the proposed algorithms are heuristics since there is no optimization on the customer order. One way to determine the optimal customer order is by exhausting all the possible options, which unfortunately is not possible in polynomial time for a residential community comprising of several hundred customers (due to NP-hardness of the problem). Thus, for any given customer order (among the huge number of possibilities), our algorithms and framework could be used to study the effectiveness of customer engagement plans and to obtain useful guidelines for designing these plans.

\subsection{Sub-Problem 1: For Shiftable Loads}
We develop a distributed algorithm in which the grid operator provides coordination among the customers. To be mindful of the security and privacy concerns of the customers, only the aggregated load profiles are exchanged between the grid operator and customers. The given aggregated load profile of the community, i.e., $\textbf{x}$, is communicated to the home controller of customer 1, which runs an algorithm developed for shiftable loads (which will be explained below). This algorithm determines the scheduling time slots for the shiftable loads of customer 1 with the objective of peak load minimization and constraint \eqref{shift_const}. The new aggregated load profile denoted by $\hat{\textbf{x}}_1$ is sent back to the grid. The grid operator communicates $\hat{\textbf{x}}_1$ to the home controller of customer 2. This process is repeated and the aggregated profile obtained from the last customer $J$ in the sequence is denoted by $\hat{\textbf{x}}_J$, which is also the final aggregated output profile $\hat{\textbf{x}}$ for sub-problem 2. 

\subsubsection{Algorithm for Shiftable Loads}
We now explain the algorithm for shiftable loads that is carried out at the home controller. The starting time of the shiftable load $i$ of customer $j$ can lie anywhere in the interval $t_{j,i}^\text{start} \in \{t_{j,i}^\text{pref}, t_{j,i}^\text{pref}+1, \ldots, t_{j,i}^\text{pref}+\Delta s_{i}^\text{max} \}$.
Thus, when a home controller $j$ receives an aggregated load profile $\hat{\textbf{x}}_{j-1}$ from the grid, the algorithm determines the best starting time for each of its shiftable load in order to minimize the peak load. Let us define a circular shift operator denoted by $\mathcal{F}(.)$, which re-arranges the entries of a vector by shifting them one unit to the right and moving the last entry to the first position\footnote{We assume that the demanded scheduling parameters of all the shiftable loads are defined in such a way that despite being delayed they can always be completed until midnight i.e., we do not allow tasks to spill over to next day.}. The scheduling algorithm for a specified sequence of shiftable devices at the home controller of customer $j$ is given as Algorithm~\ref{greedy:algo}. 
\begin{algorithm}[htb]
\caption{Algorithm For Shiftable Loads}
\label{greedy:algo}
\begin{algorithmic}[1]
\STATE{Initialize: $\textbf{p}_{j,i}^1=\textbf{p}_{j,i},\: \forall i\in I_S^j$, $\textbf{y}=\hat{\textbf{x}}_{j-1}-\sum_{i\in I_S^j} \textbf{p}_{j,i}$}
\FOR{$i \in I_S^j$} 
\FOR{$l=1:\Delta s_{i}^\text{max}$}
\STATE{$\textbf{y}^l=\textbf{y}+\textbf{p}_{j,i}^l$}
\STATE{Find the peak: $\phi_l=\max\:\textbf{y}^l$}
\STATE{$\textbf{p}_{j,i}^{l+1}=\mathcal{F}(\textbf{p}_{j,i}^l)$}
\ENDFOR
\STATE{Find the index: $l_i^*=\min_l: \{\phi_1,\ldots, \phi_{\Delta s_{i}^\text{max}}\}$}
\STATE{$t_{j,i}^\text{start}=t_{j,i}^\text{pref}+l_i^*-1$}
\STATE{$\textbf{y}=\textbf{y}^{l_i^*}$}
\ENDFOR
\end{algorithmic}
\end{algorithm}

The optimal device order, which shiftable loads should be considered for scheduling, can be determined by exhausting all the possibilities, and this requires Algorithm \ref{greedy:algo} to be executed $(I_S^j)!$ times. Let the index $m$ denote a particular order of shiftable devices. For this order we denote the final output of Algorithm \ref{greedy:algo} by $\textbf{y}_m$. Let $\hat{\phi}_m=\max\:\textbf{y}_m$ denote the peak load. Then the optimal device order denoted by $m^*$ is the one that corresponds to the minimum value in the set $\{\hat{\phi}_1,\ldots, \hat{\phi}_{(I_S^j)!}\}$ 
and the aggregated output profile $\hat{\textbf{x}}_{j}=\textbf{y}_{m^*}$ is communicated back to the grid. Typically the number of shiftable devices in each home is small (1 to 5 devices), which makes it possible to determine the optimal device order and hence the optimal output profile.

\subsection{Sub-Problem 2: For Thermostat Loads}
For thermostat loads, we again develop a distributed sequential algorithm coordinated by the grid. The grid operator communicates the aggregated load profile $\hat{\textbf{x}}$ that is obtained from the solution of sub-problem 1 to the home controller of customer 1. At the home controller we have an algorithm for thermostat loads (which will be explained below) and the load profile obtained from this algorithm denoted by $\tilde{\textbf{x}}_1$ is sent back to the grid operator for onward transmission to customer 2 and the process is repeated until we obtain the final aggregated output profile $\tilde{\textbf{x}}$.

The power consumption of thermostat load $i$ of customer $j$ at time $t$ when it is operating in state $k$ can be modeled in terms of the rated power of the appliance by the following equation,
\begin{equation}
q_{j,i}^k(t)= \frac{k-1}{K_i^j-1}\times p_{j,i}^r, \quad \forall j,k,i\in {I_T^j}
\label{power_con_model}
\end{equation}
For example, an AC that can operate in 5-states has the capability to throttle power at 0\%, 25\%, 50\%, 75\% and 100\% of the rated power. Let us define a $T\times K_i^j$ matrix $\textbf{Q}_{j,i}$, where the entries $q_{j,i}^k(t)$ of this matrix are given according to \eqref{power_con_model}. 

We also require thermal load models that can relate the output temperature obtained by an appliance when it is operated in some power throttling state. Each thermostat load has its own thermal characteristics and hence the modeling varies. Since AC and WH are the most significant thermostat loads, we discuss their thermal load models in details. We adapt a simple model, where the output temperature depends only on the current state $k$ in which the device is operated i.e., $\theta_{j,i}(\textbf{c}_{j,i}(t))=\theta_{j,i}^k(t)$.

\subsubsection{Thermal Model of AC}
The output of an AC is the room temperature that can be obtained by its operation. The room temperature variation between any two consecutive time intervals is modeled by the following equation (inspired from \cite{g14}),
\begin{equation}
 \theta_{j,\text{AC}}^k(t+1) - \theta_{j,\text{AC}}^k(t) = \Delta{t} \frac{G_j(t)}{\alpha} + \Delta{t} \frac{\hat{Z}^{k}_{j,\text{AC}}}{\alpha} w_{j,\text{AC}}(t)
 \label{Ti+1:hvac}
 \end{equation}
In this model, $G_j(t)$ is the heat gain rate of the house of customer $j$, which depends on heat gain coefficients of the walls, windows, roof, solar radiation, people and air change rate of the AC, inside and outside temperature difference, etc., ($G_j(t)$ is independent of state $k$), $\alpha$ is the energy required for a unit degree rise in room temperature and $\hat{Z}^{k}_{j,\text{AC}}$ is the cooling capacity of the AC when it is operating in state $k$. Cooling capacity of an AC is specified by the manufactures in kW, BTU/hr or Tons. The cooling capacity is a function of the power state in which AC is operated and can be modeled as, 
\begin{equation}
\hat{Z}^{k}_{j,\text{AC}} = \textit{EER} \times q_{j,i}^k(t) 
\end{equation}
where $EER$ is the Energy Efficiency Ratio, which is typically defined as the ratio of cooling capacity given in BTU/hr to the power input in Watts. The higher the EER rating, the greater is the performance. The US national appliance standards dictate all AC loads to have a minimum value of $EER \geq 8.0$ \cite{energygov}. For example, an AC with a cooling capacity of 1 Ton (equivalent to 12000 BTU/hr or 3.516 kW) with an $EER$ of 8.0 will consume 1.5 kW power.

\subsubsection{Thermal Model of WH}
Hot water temperature obtained from operating the WH for one time slot in any given power state $k$ can be modeled by the following equation (inspired from \cite{g14}),  
\begin{equation}
\begin{split}
\theta_{j,\text{WH}}^k(t+1)& =  \frac{ \theta_{j,\text{WH}}^k(t) (V_\text{tank}-fr(t) . \Delta{t})}{V_\text{tank}} + \frac{\theta_\text{inlet} . fr(t) . \Delta{t}}{V_\text{tank}}\\ 
& +  \bigg[ q_{j,i}^k(t) - \frac{A_\text{tank} . (\theta_{j,\text{WH}}^k(t) - 
\theta_{a}(t))}{R_\text{tank}}\bigg] . \frac{\Delta{t}}{V_\text{tank}} 
\end{split}
\label{wh_eq}
\end{equation}
In this model, water temperature in the next time slot $t+1$, by operating WH in power state $k$ (which consumes $q_{j,i}^k(t)$ amount of power) depends on water temperature at the start of the time interval, temperature of the inlet water ($\theta_\text{inlet}$), current room temperature ($\theta_a(t)$), water flow rate ($fr(t)$) during time slot $t$, volume of the tank ($V_\text{tank}$), tank surface area ($A_\text{tank}$) and the heat resistance of the water tank ($R_\text{tank}$). In this model, there are three terms and some necessary conversions might be required to make the units of all these terms consistent.

    
\subsubsection{Algorithm for Thermostat Loads} The home controller of user $j$ receives the aggregated load profile denoted by $\tilde{\textbf{x}}_{j-1}$ from the grid. The objective of Algorithm \ref{therm:alg} again is to reduce the peak load, while respecting constraints \eqref{oper_const1}, \eqref{oper_const2} and \eqref{temp_const} as laid out in the customer's terms of engagement with the grid. There are two inconvenience dimensions (severity and duration) for each thermostat load. Thermostat loads are considered in a sequential order. Each thermostat load can be denied operation at its full rated power for a maximum of $t_i^\text{dur}=\frac{\Delta d_{i}^\text{max}}{\Delta t}$ number of time slots in its demanded operation interval. Let $\textbf{b}_{j,i}$ denote the set of time indexes during which customer $j$ has demanded thermostat load $i$. For each thermostat load, the algorithm determines $t_i^\text{dur}$ local peaks in the interval $\textbf{b}_{j,i}$. Let, $t_{j,i}^\text{max}=\max_{\textbf{b}_{j,i}}\: \textbf{h}$ and $t_{j,i}^{\text{max}-n}, n=1,\ldots, t_i^\text{dur}+1$ represent the time index corresponding to $n$-th largest local peak. The time index vector representing the local peaks arranged in descending order is then denoted by $\textbf{t}_{j,i}^\text{peak}=[t_{j,i}^\text{max}, \: t_{j,i}^{\text{max}-1}, \: \ldots , t_{j,i}^{\text{max}-t_i^{\text{dur}}-1} ]$. The objective is to operate the thermostat load $i$ of customer $j$ in its lowest possible state at time index $t_{j,i}^\text{max}$, followed by $t_{j,i}^{\text{max}-1}$ until $t_{j,i}^{\text{max}-t_i^\text{dur}-1}$, without violating the inconvenience severity constraint anywhere in the demanded interval $\textbf{b}_{j,i}$. The algorithm starts by switching OFF the thermostat load $i$ of customer $j$ at all the time indexes in vector $\textbf{t}_{j,i}^\text{peak}$ (i.e., $c_{j,i}^1(t)=1,\: \forall t \in \textbf{t}_{j,i}^\text{peak}$). Once the operational states are fixed at the desired time indexes, the algorithm determines a sequence of output temperatures denoted by $\bar{\theta}_{j,i}(t), \forall t\in \textbf{b}_{j,i}$. If the inconvenience severity constraint is not violated anywhere in the demanded time interval, the algorithm will terminate. Otherwise, the algorithm will increase the operational state of the thermostat load $i$ at the time index $t_{j,i}^{\text{max}-t_i^\text{dur}-1}$ (i.e., $c_{j,i}^1(t_{j,i}^{\text{max}-t_i^\text{dur}-1})=0$ and $c_{j,i}^2(t_{j,i}^{\text{max}-t_i^\text{dur}-1})=1$) and recomputes the output temperature sequence. This process is repeated until the inconvenience severity constraint is satisfied everywhere in demanded operational interval of the customer. This algorithm ensures that at the highest local peak points, the thermostat load is either switched OFF or it is operating in the lowest possible power state. Since least preference is given to the lowest local peak points, the thermostat load might operate at the full rated power at these points. Thus, for some customers, the algorithm will achieve the inconvenience duration constraint with strict inequality, in order to satisfy inconvenience severity constraint in all of its demanded time slots\footnote{The worst case complexity in terms of number of iterations is $\sum_{i\in I_T^j} (K_i^j)!$. For example, if we consider two thermostat loads per customer each having 5-states, the worst case complexity of Algorithm \ref{therm:alg} is 240 iterations. We show in the simulations, that increasing the power states beyond three is not much beneficial for peak load reduction. The number of thermostat loads as well as the power throttling states of devices are generally low, therefore, this complexity is manageable. }.   
\begin{algorithm}[htb]
\caption{Algorithm for Thermostat Loads}
\label{therm:alg}
Initialize: $\textbf{h}=\tilde{\textbf{x}}_{j-1}-\sum_{i\in I_T^j} \textbf{p}_{j,i}$, $\textbf{C}_{j,i}=\textbf{0}, \forall i \in I_T^j$
\begin{algorithmic}[1]
\FOR{$i \in I_T^j$ }
\STATE{Determine and arrange in descending order the time indexes of $t_i^\text{dur}$ local peaks in the aggregated profile $\textbf{h}$ in the time interval $\textbf{b}_{j,i}$ denoted by: $\textbf{t}_{j,i}^\text{peak}=[t_{j,i}^\text{max} \: t_{j,i}^{\text{max}-1} \: \ldots t_{j,i}^{\text{max}-t_i^{\text{dur}}-1} ]$}
\STATE{Initialize: $k=0$, $t+a=t_{j,i}^{\text{max}-t_i^{\text{dur}}-a}$, $a=1$, $c_{j,i}^1(t)=1,\: \forall t \in \textbf{t}_{j,i}^\text{peak}$, $\bar{\theta}_{j,i}(t)=0,\forall t\in \textbf{b}_{j,i}$}
\WHILE{$|\bar{\theta}_{j,i}(t)-\hat{\theta}_{j,i}|>\Delta \theta_{j,i}$ for any $t \in \textbf{b}_{j,i}$}
\STATE{$c_{j,i}^{k}(t+a)=0$}
\STATE{Increment: $k=k+1$}
\STATE{$c_{j,i}^{k+1}(t+a)=1$}
\STATE{Determine $\bar{\theta}_{j,i}$ using the thermal load model}
\IF{$k=K_i^j$}
\STATE{$k=0$, \: $a=a+1$}
\ENDIF
\ENDWHILE
\STATE{$\textbf{p}_{j,i}=(\textbf{Q}_{j,i} \odot \textbf{C}_{j,i}) \textbf{1}_{K_i^j}$ (Note: $\odot$ denotes element-wise matrix multiplication)}
\STATE{Update the aggregated load profile: $\textbf{h}=\textbf{h}+\textbf{p}_{j,i}$}
\ENDFOR
\end{algorithmic}
\end{algorithm}

This algorithm again requires optimization over all devices orders, since considering thermostat loads in a different sequence can result in a different amount of peak load reduction. This can be done by executing Algorithm \ref{therm:alg} for $(I_T^j)!$ times at home controller of customer $j$. The number of thermostat loads in every home is limited, therefore, this optimization step does not result in significant increase in the complexity. For instance, if we assume that every customer has only two thermostat loads (AC and WH), then there are only two device orders: AC followed by WH or WH followed by AC, and Algorithm \ref{therm:alg} will only be executed twice.

\begin{table}[htb]
\label{pr_datax}
\caption{Appliance power rating and usage patterns}
\centering
\begin{tabular} {|c|  c| c| c| }
\hline

\multirow{1}* {Flexible} & \multirow{1}* {Power Rating} & \multirow{1}* {Operation} & \multirow{1}* {Usage} \\
{Load} & {(kW)} & {Duration} & {Pattern}  \\
\hline

\multirow{2}* {AC} & \multirow{2}* {5} & \multirow{2}* {4 hours} & \multirow{1}* {Consecutive} \\
 &  &  & {operation}  \\
\hline

\multirow{2}* {WH} & \multirow{2}* {2.5} & \multirow{2}* {3-6 hours} & \multirow{1}* {Two or three separate} \\
 &  &  & {instances of 1-2 hours}  \\
\hline

\multirow{2}* {CD} & \multirow{2}* {3.1} & \multirow{2}* {2 hours} & \multirow{1}* {Consecutive} \\
 &  &  & {operation}  \\
\hline

\multirow{2}* {DW} & \multirow{2}* {1.8} & \multirow{2}* {1.5 hours } & \multirow{1}* {Consecutive} \\
 &  &  & {operation}  \\
\hline

\end{tabular}
\end{table}

\section{Simulation Results}
\begin{figure}[htb]
\centering
\includegraphics[width=0.4\textwidth,height=.13\textheight]{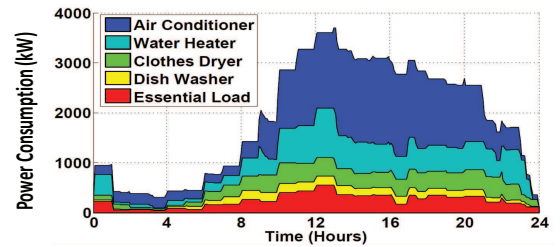}
\caption{Aggregate load profile of the residential community } 
\label{fig:agg}
\end{figure}

We consider a residential community comprising of 1000 homes. Each customer is assumed to have two thermostat loads i.e., AC and WH and two shiftable loads i.e., CD and DW. All other devices contribute towards an essential base load. The average daily household energy consumption is assumed to be about 41 kWh, which corresponds to typical household energy consumption in many US states like Louisiana, Tennessee, Alabama etc. The appliance usage and the power consumption pattern of each device is given in Table II, which are generated using realistic assumptions as given in \cite{energy1,energy_pattern} and also to match a load curve shape from the RELOAD database on a typical summer day. The RELOAD database is an industry accepted database of load curve shapes and is used by the electricity module of the National Energy Modeling System (NEMS) and several authors for their studies \cite{reload_1}, \cite{reload_2}, \cite{g14}. The thermostat set points of AC and WH are uniformly distributed random variables in the interval [68$^o$F,76$^o$F] and [104$^o$F,120$^o$F]. The resulting aggregated load profile of the community is shown in Fig. \ref{fig:agg}, with a peak load value of 3701 kW. In the simulations, we assume $K_i^j=K, \forall j, i\in I_T^j$ and evaluate the impact of 2-state, 3-state, and 5-state models for AC and WH. The EER value of AC load is assumed to be 10. We run the simulations in MATLAB and divide the total time duration i.e., 24 hours into $T$=288 equal interval time segments ($\Delta t$ = 5 mins).

\begin{figure}[htb]
\centering
\includegraphics[width=0.4\textwidth,height=.15\textheight]{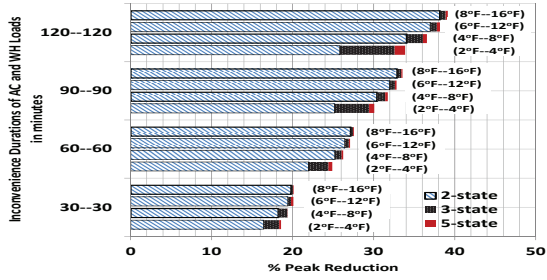}
\caption{Percentage peak reduction achieved by CDPs. Y-axis shows the maximum inconvenience durations of AC and WH loads in the format: $\Delta d_\text{AC}^\text{max}--\Delta d_\text{WH}^\text{max}$. The maximum temperature deviations of AC and WH loads are shown along the bars in the format: ($\Delta \theta_\text{AC}^\text{max}--\Delta \theta_\text{WH}^\text{max}$). }
\label{fig:peak_fix}
\end{figure} 

\begin{figure}[htb]
\centering
\includegraphics[width=0.4\textwidth,height=.15\textheight]{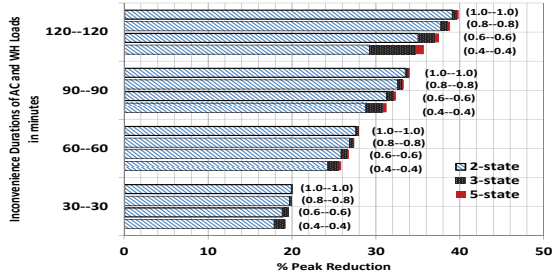}
\caption{Percentage peak reduction achieved by PDPs. Y-axis shows the maximum inconvenience durations of AC and WH loads in the format: $\Delta d_\text{AC}^\text{max}--\Delta d_\text{WH}^\text{max}$. The scaling factors of AC and WH loads are shown along the bars in the format: ($\beta_\text{AC}--\beta_\text{WH}$).  }
\label{fig:peak_var}
\end{figure}

We evaluate the effectiveness of CDPs and PDPs and the impact of increasing the power throttling states on peak load reduction in Figs. \ref{fig:peak_fix} and \ref{fig:peak_var} respectively. The common plan parameters in both these figures are: $\Delta s_\text{CD}^\text{max}=0$ mins and $\Delta s_\text{DW}^\text{max}=0$ mins, $\theta_\text{AC}^\text{ref}=80^o$F, $\theta_\text{WH}^\text{ref}=96^o$F. In Fig. \ref{fig:peak_fix}, we vary the parameters, which define the temperature deviations and inconvenience durations of AC and WH loads for CDPs. While in Fig. \ref{fig:peak_var}, we vary the scaling factors and inconvenience durations of AC and WH loads for PDPs. Based on these two figures, we have the following observations.
\begin{itemize}
	\item Increasing the inconvenience durations of AC and WH loads, the temperature deviations in CDPs, the scaling factors in PDPs, and the number of states of AC and WH loads increase the peak load reduction. 
		\item When inconvenience durations of AC and WH loads in CDPs and PDPs are fixed, we observe a marginal peak load reduction with increasing temperature deviations in CDPs and scaling factors in PDPs (also termed as diminishing returns). This is due to the saturation of the inconvenience severity dimension of the plan. Similarly, when temperature deviations of AC and WH loads in CDPs and scaling factors in PDPs are fixed, we again observe diminishing returns when the inconvenience durations of loads are increased. This is due to the saturation of the inconvenience duration dimension of the plan.
		\item The inconvenience severity dimension saturates at a faster rate compared to the inconvenience duration dimension of the plan for both the CDPs and PDPs.
	\item For both CDPs and PDPs, diminishing returns are observed with an increase in the number of power throttling states. 	
	\item There is a tradeoff between the returns on the number of states and inconvenience severity dimension of the plan. Plans offering less inconvenience severity (less temperature deviations for CDPs and less scaling factors for PDPs) exhibit more gains with the increase in the number of power states. On the other hand, plans with high inconvenience severity, do not offer any significant gains with the increase in the number of states. In high inconvenience severity plans, intermediate states are generally not required even if they are available, since operating the loads in OFF states save more power without violating the inconvenience severity constraints. 
\end{itemize}

\begin{figure}[htb]
\centering
\includegraphics[width=0.48\textwidth,height=.28\textheight]{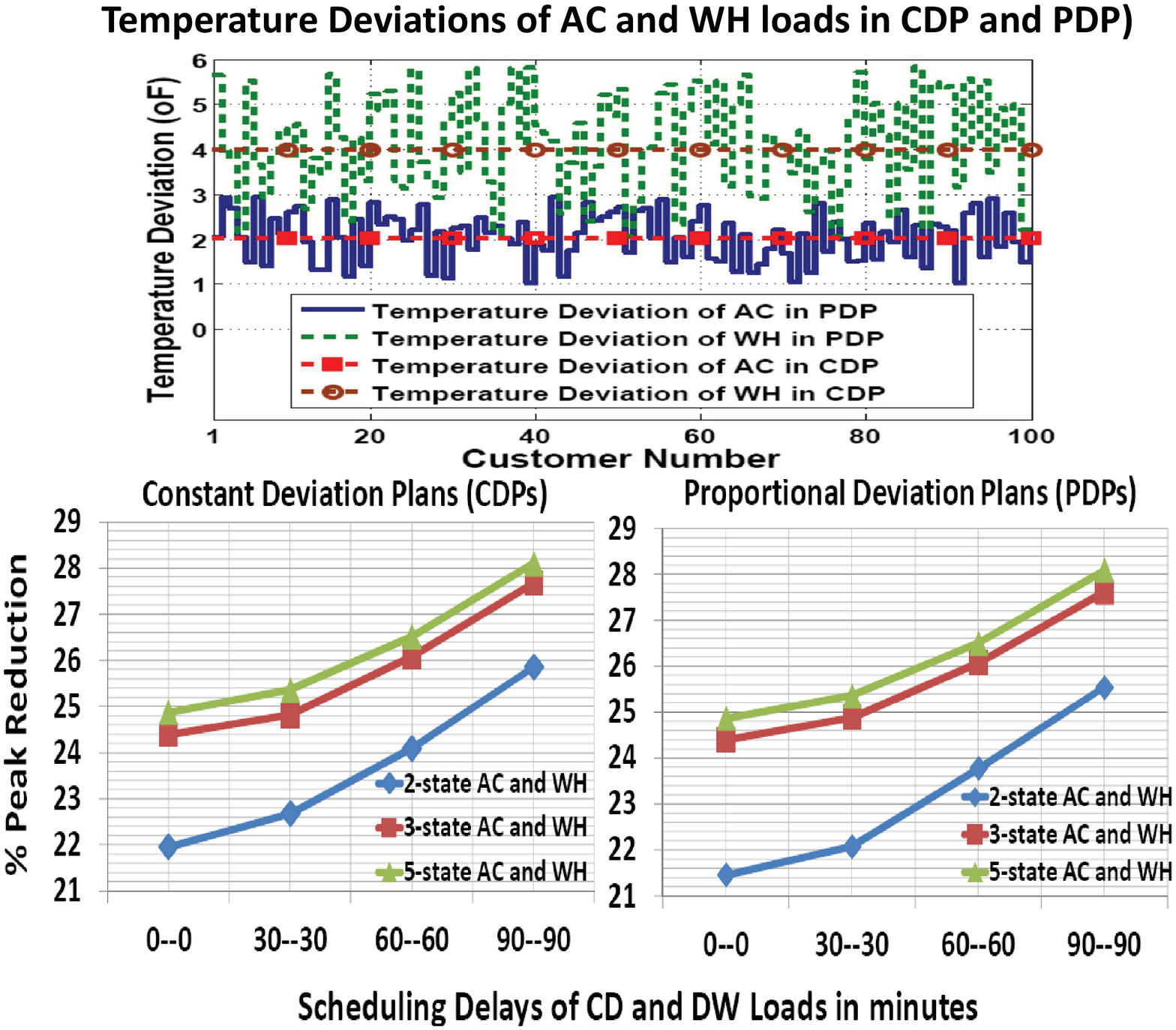}
\caption{The top graph shows the temperature deviations experienced by the AC and WH loads of 100 customers in the community in CDP and PDP. The bottom graphs show the percentage peak reduction for different scheduling delays for CD and DW loads in CDPs and PDPs. X-axis of the bottom graph shows the maximum scheduling delays of CD and DW loads in minutes, in the format:  $\Delta s_\text{CD}^\text{max}--\Delta s_\text{DW}^\text{max}$.  }
\label{fig:figdel}
\end{figure}

In Fig. \ref{fig:figdel}, we plot the impact of scheduling delays of CD and DW loads on the peak load reduction. The CDP plan parameters are: AC load (60 min, 2$^o$F, 80$^o$F), WH load (60 min, 4$^o$F, 96$^o$F), while PDP parameters are: AC load (60 min, 0.25, 80$^o$F), WH load (60 min, 0.25, 96$^o$F). The temperature deviations experienced by the AC and WH loads of different customers are different in PDP but constant in CDP, which can be visualized from the top graph in this figure. In these simulations, the choice of $\beta_\text{AC}=\beta_\text{WH}=0.25$ in PDPs, results in an average temperature deviation of 2.04$^o$F for AC load and 4$^o$F for WH load, which match the temperature deviations of 2$^o$F for AC and 4$^o$F for WH in CDPs. From the bottom graphs, we can observe that as we increase the scheduling delays, the amount of peak load reduction and the rate of peak load reduction both increase. Furthermore, PDPs, while offering more fair temperature deviations (inversely proportional to thermostat set points), also achieve almost identical peak load reduction that is obtained by the CDPs.  

\begin{table*}[htb]
\label{tab_FSP}
\renewcommand{\arraystretch}{1.6}
\renewcommand{\tabcolsep}{7pt}
\caption{This table shows the impact of reference temperatures on the number of eligible loads, average temperature deviations over the demanded interval and \% peak reduction for CDPs and PDPs.  }
\centering
\begin{tabular}{|c|c|c|c|c|c|c|c|c|c|c|c|c|c|}
\hline

    \multicolumn{2}{|c|}{\multirow{2}{*}{Ref. Temp.}} &
		\multicolumn{2}{c|}{\multirow{2}{*}{No. of eligible loads}} & 
    \multicolumn{4}{c|}{Average Temperature Deviations } & 
		\multicolumn{6}{c|}{\% Peak Reduction } \\
\cline{5-14}
  \multicolumn{2}{|c|} {} & \multicolumn{2}{c|} {} & \multicolumn{2}{c|}{$\theta_\text{AC}^\text{ave}$} &  \multicolumn{2}{c|} {$\theta_\text{WH}^\text{ave}$} & \multicolumn{2}{c|}{2-state} &  \multicolumn{2}{c|} {3-state} &  \multicolumn{2}{c|} {5-state} \\  \hline
	$\theta_\text{AC}^\text{ref}$ & $\theta_\text{WH}^\text{ref}$ & $N_\text{AC}$ & $N_\text{WH}$ & CDP & PDP & CDP & PDP & CDP & PDP & CDP & PDP & CDP & PDP  \\  \hline


72$^o$F & 112$^o$F & 550 & 509 & 1.51$^o$F & 0.5$^o$F & 2.97$^o$F & 1.0$^o$F & 14.98 & 6.13 & 22.66 & 20.71 & 24.49 & 23.45 \\ \hline

74$^o$F & 108$^o$F & 766 & 752 & 1.71$^o$F & 0.79$^o$F & 3.32$^o$F & 1.50$^o$F & 19.36 & 8.07 & 23.95 & 21.27 & 25.4 & 24.28  \\ \hline

76$^o$F & 104$^o$F &  1000 &  1000 & 1.76$^o$F & 1.04 & 3.48 & 2.0 & 22.84 & 13.55 & 25.45 & 22.58 & 26.22 & 25.03 \\ \hline

78$^o$F & 100$^o$F & 1000 & 1000 & 2$^o$F & 1.54$^o$F & 4$^o$F & 3$^o$F & 24.1 & 21.22 & 26.09 & 24.66 & 26.52 & 25.91  \\ \hline

80$^o$F & 96$^o$F & 1000 & 1000 & 2$^o$F & 2.04$^o$F & 4$^o$F & 4$^o$F & 24.1 & 23.78 & 26.09 & 26.08 & 26.52 & 26.51 \\ \hline

\end{tabular}

\end{table*}

Finally, we study the impact of reference temperature on peak reduction for CDPs and PDPs and the results are presented in Table III. Common Parameters of CDPs and PDPs are $\Delta s_\text{CD}^\text{max}=\Delta s_\text{DW}^\text{max}=\Delta d_\text{AC}^\text{max}=\Delta d_\text{WH}^\text{max}=60$ min. CDP parameters: $\Delta \theta_\text{AC}^\text{max}=2^o$F, $\Delta \theta_\text{WH}^\text{max}=4^o$F. PDP parameters: $\beta_\text{AC}=\beta_\text{WH}=0.25$. In this table, $N_\text{AC}$ and $N_\text{WH}$ respectively denotes the number of AC and WH loads with non-zero inconvenience severity, while $\theta_\text{AC}^\text{ave}$ and $\theta_\text{WH}^\text{ave}$ respectively denotes the average temperature deviations experienced by $N_\text{AC}$ and $N_\text{WH}$ loads. Different values of reference temperatures result in different values of $N_\text{AC}$ and $N_\text{WH}$. As we increase the reference temperature for AC and decrease that of WH, more peak load reduction is observed for CDPs and PDPs, since more devices are controlled. We can observe another tradeoff between the number of states and reference temperatures (which impact $N_\text{AC}$ and $N_\text{WH}$) especially for PDPs. When the reference temperatures result in less number of ACs and WHs experiencing inconvenience, providing an additional power throttling state can provide very high performance gains compared to the 2-state (basic ON/OFF switching) models. We can also observe a similar trend (but in lesser magnitude) for CDPs. Furthermore, when $\theta_\text{AC}^\text{ref} \geq \hat{\theta}_{j,\text{AC}}+\Delta \theta_\text{AC}^\text{max}, \forall j$ and $\theta_\text{WH}^\text{ref} \leq \hat{\theta}_{j,\text{WH}}-\Delta \theta_\text{WH}^\text{max}, \forall j$, then increasing $\theta_\text{AC}^\text{ref}$ and decreasing $\theta_\text{WH}^\text{ref}$ has no further impact on the peak load reduction in CDPs (since there is no further impact on the number and temperature deviations, as evident from the last two rows of Table III). On the other hand, in PDPs, increasing $\theta_\text{AC}^\text{ref}$ and decreasing $\theta_\text{WH}^\text{ref}$ always increase the amount of peak load reduction (since temperature deviations also increase). 

 
\section{Conclusions and Future Research Directions}
In this paper, we have designed two types of customer engagement plan, namely CDP and PDP, which describe the key inconvenience parameters of flexible loads in terms of scheduling delays and temperature deviations so as to make the customers easily understand the inconvenience caused by these plans. To facilitate the grid operator in determining the effectiveness of such plans on peak load reduction, we have developed appropriate DRP algorithms by modeling the thermostat loads as power throttling devices. Despite the suboptimality (due to NP nature of the optimization problem), the proposed algorithms have been shown to be able to provide very clear insights into the design of customer engagement plans. Through simulations, we have determined that increasing temperature deviations result in diminishing returns. We have also observed diminishing returns as the number of power throttling states increase. In particular, more peak load reduction occurs when the number of power throttling states are increased from 2 to 3 for those DRP plans with less number of thermostat loads. The temperature deviation from thermostat set point cannot exceed the given limit in all the time slots, therefore, with two states, the only option to avoid exceeding the temperature deviation limit is to turn ON the appliance (consuming full power). On the other hand, the third state provides more flexibility and we can avoid exceeding the temperature deviation limit by operating the device at 50\% power (hence resulting in 50\% power reduction) compared to the two state model. 

We also have some recommendations for the design of customer engagement plans. The grid operator should generally design plans with low to moderate inconvenience severity. In a community with thermostat loads having more fine control (three or more states), the plans that offer low inconvenience severities are more beneficial. On the other hand, for a residential community with only ON/OFF power control for thermostat loads, more peak reduction can be achieved by designing plans with high inconvenience duration of thermostat loads and scheduling delay of shiftable loads. PDPs can be offered to the customers of a residential with highly variable thermostat set point preferences, while CDPs can be offered to the customers with similar thermostat set point preferences. The grid operator can use the reference temperatures in the plans to control the number of eligible devices. The value of reference temperatures can also be adjusted by the grid operator depending on the weather patterns, geographical location, user preferences, surveys etc. 

These results can be seen as useful guidelines in proposing financial rewards and mutually beneficial incentives, which is an interesting future work. Extending the framework to include more load categories e.g., electrical vehicles are another interesting research direction. Modeling the transient behavior and reducing the additional power consumption due to frequent switching of thermostat appliances from one power throttling state to another might also be an interesting future work. Finally, grid operator may also introduce multiple plans, each with different parameters, to cater for the preference of different customers. Managing and designing various plans also appears to be an interesting future work.

\end{document}